\documentclass{ifacconf}

\usepackage{graphicx}      % include this line if your document contains figures
\usepackage{natbib}        % required for bibliography

\usepackage{amsmath,graphicx}
\usepackage{tikz}
\tikzset{%
	block/.style    = {draw, thick, rectangle, minimum height = 4em,
		minimum width = 4em},
	sum/.style      = {draw, circle, node distance = 3cm}, % Adder
	input/.style    = {coordinate}, % Input
	output/.style   = {coordinate} % Output
}
\usepackage[framemethod=TikZ]{mdframed}
\usepackage{lipsum}
\newcommand\blfootnote[1]{%
  \begingroup
  \renewcommand\thefootnote{}\footnote{#1}%
  \addtocounter{footnote}{-1}%
  \endgroup
}
\mdfdefinestyle{MyFrame}{%
	linecolor=blue,
	outerlinewidth=2pt,
	roundcorner=20pt,
	innertopmargin=\baselineskip,
	innerbottommargin=\baselineskip,
	innerrightmargin=20pt,
	innerleftmargin=20pt,
	backgroundcolor=gray!50!white}
\begin{document}
\begin{frontmatter}

\title{ARX Model Identification using Generalized Spectral Decomposition} 

% \thanks[footnoteinfo]{Sponsor and financial support acknowledgment goes here. Paper titles should be written in uppercase and lowercase letters, not all uppercase.}

\author[First]{Deepak Maurya} 
\author[Second]{Arun K. Tangirala} 
\author[Third]{ Shankar Narasimhan}

\address[First]{Department of Computer Science, 
   Indian Institute of Technology Madras, (e-mail: maurya@cse.iitm.ac.in)}
\address[Second]{Department of Chemical Engineering, 
   Indian Institute of Technology Madras, (e-mail: arunkt@iitm.ac.in)}
\address[Third]{Department of Chemical Engineering, 
   Indian Institute of Technology Madras, (e-mail: naras@iitm.ac.in)}

\begin{abstract}                % Abstract of not more than 250 words.
This article is concerned with the identification of autoregressive with exogenous inputs (ARX) models. Most of the existing approaches like prediction error minimization and state-space framework are widely accepted and utilized for the estimation of ARX models but are known to deliver unbiased and consistent parameter estimates for a correctly supplied guess of input-output orders and delay. 

In this paper, we propose a novel automated framework which recovers orders, delay, output noise distribution along with parameter estimates. The primary tool utilized in the proposed framework is generalized spectral decomposition. The proposed algorithm systematically estimates all the parameters in two steps. The first step utilizes estimates of the order by examining the generalized eigenvalues, and the second step estimates the parameter from the generalized eigenvectors. Simulation studies are presented to demonstrate the efficacy of the proposed method and are observed to deliver consistent estimates even at low signal to noise ratio (SNR). 
\end{abstract}

\begin{keyword}
System identification, eigenvalue problem, ARX model, Linear systems, Order determination 
\end{keyword}

\end{frontmatter}
%===============================================================================

\section{INTRODUCTION}
\blfootnote{\textsuperscript{\textcopyright} 20XX the authors. This work has been accepted to IFAC for publication under a Creative Commons Licence CC-BY-NC-ND}
\label{sec:intro}
System identification is a broad field that primarily deals with the study of parameter estimation of a formulated model \citep{book:ljung}. It has been an active area of research due to its numerous applications in various domains like signal processing  \citep{book:oppenheim1999discrete}, control and fault detection \citep{sepulchre2012constructive,soton368264} tasks.

An ideal system identification algorithm is expected to deliver efficient and consistent parameter estimates from noisy signals in an automated manner requiring minimal user intervention. Accurate  knowledge of these model parameters improves the quality of several tasks like spectral analysis, filtering \citep{book:haykin2008adaptive} and controller design \citep{paper:elliott1993active}. These basic functions can be used in various specific applications like room acoustics \citep{paper:xue2017frequency}, speaker localization \citep{paper:doclo2003robust}, room geometry estimation \citep{paper:moore2013room}.

\begin{figure}[htb]
	\begin{minipage}[b]{1.0\linewidth}
		\centering
		\centerline{\includegraphics[width=5cm]{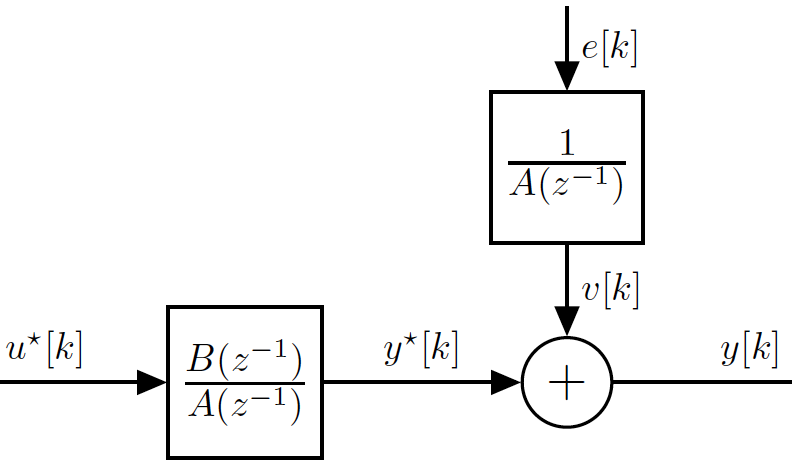}}
		%  \vspace{2.0cm}
	\end{minipage}
	\caption{ARX Model Architecture}
	\label{fig:arx_process}
\end{figure}

% One of the classical problems in signal processing is adaptive noise cancellation (ANC) \cite{paper:glover1977adaptive,paper:widrow1975adaptive}. In ANC literature as seen from Figure 1 of \cite{paper:friedlander1982system}, $y[k]$ is referred as ``primary'' input and $u[k]$ as ``reference'' input .  In system identification literature, $y[k]$ is referred as measured output and $u[k]$ as input. The first milestone in this problem is estimation of the transfer function or filter relating $y[k]$ and $u[k]$. 
% \cite{paper:friedlander1982system} and \cite{paper:friedlander1982_1} have considered Box-Jenkins (BJ) and ARMAX model among $y[k]$ and $u[k]$ respectively.
In this work, we consider the problem of model estimation of a general ARX model for single input and single output (SISO) systems as shown in Figure \ref{fig:arx_process}. 

As this is a fundamental problem, abundant literature exists in the field system identification \citep{book:ljung, georgiou1992identification}. Most of the existing algorithms like prediction error minimization (PEM),  state-space framework, recursive ARX estimation \citep{book:akt_sysid} deliver an unbiased estimate of the parameter for correctly supplied model structure. We refer to input-output orders and delays as the model structure in this paper.  

In the absence of a priori information of such critical parameters, the user is forced to try multiple guesses of order and delay until the convergence to satisfactory results \citep{book:akt_sysid}. Most of the existing approaches for identification of the right model structure can be clustered in three categories based on the stage of estimation:
\begin{enumerate}
    \item Pre-Estimation: As the name suggests, the user has or acquires knowledge about model structure before the estimation of the model. Various  non-parametric methods like step-response or impulse-response analysis, information metrics like Akaike Information Criteria (AIC) \citep{book:ljung,paper:beheshti2005new,paper:wax1988order} and state-space methods \citep{paper:morikawa1984system} and frequency domain approaches like Bode plots  \citep{book:akt_sysid} can be utilized for the estimation of model structure.
    \item During-Estimation:  This class of methods tries to estimate the model structure along with model parameters. The core idea is to utilize regularization or sparsity constraints in compressed sensing techniques for model structure determination \citep{paper:satheesh_CS}.
    \item Post-Estimation:  In these approaches, a certain model structure is assumed, and its correctness is checked after estimation. This class of methods includes approaches like recursively estimating ARX models, checking under parameterization, or over parameterization of the model through residual analysis or hypothesis testing of estimated parameters \citep{book:akt_sysid}. 
\end{enumerate}

The user is compelled to try most of these approaches practically in an iterative manner. For example, if the model structure is found to be unsatisfactory at the post-estimation stage, the user is bound to try other combinations of input-output order and delay from the beginning of the model estimation stage. Some of these approaches are heuristic, which operates under restrictive assumptions and lacks theoretical rigor. 

The approaches mentioned above can also be classified into parametric and non-parametric methods for model structure determination.  The prime issue with these approaches is that they operate for a given model structure and parameter estimation in an uncoupled manner. This turns out to be the prime reason for forcing the user to try multiple combinations of input-output order and delay.   

In this work, we consider the challenging problem of jointly estimating order and model parameters without any availability of a priori system information. The proposed framework provides the freedom to estimate order and model parameters jointly. With a slight abuse of terminology, this can be seen as an attempt to couple the parametric and non-parametric methods for model structure determination. The novelty of this work lies in the proposal of an automated algorithm that estimates model order, delay, and parameters without any prior knowledge on the system. 

The proposed algorithm utilizes the generalized spectral decomposition framework and is heavily inspired from \cite{paper:dycops_dipca}. Spectral decomposition framework has been gaining recent attention for system identification. Some of the closely related works on slightly different problems are by \cite{vermeersch2019globally,de2019least,zheng2018positive}.  

The proposed algorithm can be seen as comprising two major steps. The first step jointly estimates the equation order and noise distribution properties by analyzing the generalized eigenvalues of the covariance matrix for the lagged input-output data. In this paper, we refer autocovariance function (ACVF) of the noise $v[k]$ in Figure \ref{fig:arx_process} as noise distribution properties. This step requires an initial guess for the ACVF of noise, which is assumed to be not available a priori. So, we estimate the ACVF iteratively. 

The second step involves estimating the parameters of the linear difference equation relating to input and output data. In the end, the proposed algorithm systematically recovers input-output orders, delays, parameters of a difference equation, and ACVF of noise in an automated manner with minimal user intervention.

The rest of the paper is organized as follows. In Section \ref{sec:foundations}, we describe the full estimation problem formally and manifests the use of principal component analysis (PCA) and its variants in system identification. Section \ref{sec:prop} starts by presenting the equivalence of generalized spectral decomposition and existing algorithms working in the PCA framework. It also consists of the proposed algorithm under various assumptions. Most of the key ideas are demonstrated using a simple second-order system case study. Section \ref{sec:simulation} contains simulation studies that elucidate the merits of working with the proposed algorithm. Concluding remarks and directions for future work are addressed in Section \ref{sec:conc}.

% grammarly check done till here 
% --------------- foundation s -------------------------------
\section{Foundations}
\label{sec:foundations}
A general parametric deterministic linear time-invariant (LTI) dynamic model among input ($u^{\star}$) - output ($y^{\star}$) variable is described by
 \begin{align} 
	y^{\star}[k] + \sum_{i=1}^{n_y}a_{i}y^{\star}[k-i] = \sum_{j=D}^{n_u}b_ju^{\star}[k-j] \label{eq:sisodynamic}
	\end{align}
where $n_y$ and $n_u$ are output and input order, respectively and $D$ is the input-output delay. We define the term equation order, $\eta = \max(n_y,n_u)$. The problem statement is to estimate the coefficients $\{a_{i}\}_{i=1}^{n_y}$, $\{b_{j}\}_{j=D}^{n_u}$ along with $n_y$, $n_u$ and $D$ from $N$ noisy samples  of output $y^{\star}[k]$  denoted by $y[k]$ and noise-free input $u^{\star}[k]$ shown below 
 \begin{subequations}
		\begin{align}
		y[k]	&= y^{\star}[k] + v[k] \label{eq:yk_noise} \\
		v[k] &+ \sum_{i=1}^{n_y}a_{i}v[k-i] = e[k], \qquad e[k] \sim \mathcal{N}(0,\sigma_e^2) \label{eq:arx_noise}
		\end{align}
		\label{eq:dgp}
\end{subequations}
The measured output signal , denoted by $y[k]$ has ARX model structure with noise-free input $u^{\star}[k]$ as shown in Figure \ref{fig:arx_process}, where $A(z^{-1}) = 1 + \sum_{i=1}^{n_y}a_{i}z^{-1} $ and $B(z^{-1}) = \sum_{j=D}^{n_u}b_jz^{-j}$. 

\subsection{PCA \&  Dynamic PCA}
In this section, we briefly discuss the application of principal component analysis (PCA) and its variants in the context of system identification. % Further we also discuss the extension of PCA to dynamic models referred as DPCA. 
The model in Eqn. \eqref{eq:sisodynamic} can viewed as 
\vspace{-2mm} 
\begin{align}
\mathbf{\theta}^T \mathbf{x}[k] = 0
\label{eq:nullsp}
\end{align}
\vskip -5mm
{\footnotesize \begin{subequations}
		\begin{align}
		\mathbf{\theta} =  \begin{bmatrix}
		1 & a_1 & a_2 & \ldots & a_{n_y} & -b_D & -b_{D+1} & \ldots -b_{n_u} 
		\end{bmatrix}^T\\ 
		\mathbf{x}[k]  = \begin{bmatrix}
		y^{\star}[k]  & \ldots & y^{\star}[k - n_y] & u^{\star}[k-D]  \ldots u^{\star}[k-n_u]
		\end{bmatrix}^T
		\end{align} 
\end{subequations}}
Equation \ref{eq:nullsp} can be interpreted as a constrained model where the parameter vector $\mathbf{\theta}$ is orthogonal to $\mathbf{x[k]}$ for $k = 1,2,\ldots,N$. It is well established that basis for an orthogonal space can be obtained from singular vectors using singular value decomposition (SVD)  or equivalently PCA \citep{book:joliffe,paper:rao1964pca} of $\mathbf{X}$ , where  $\mathbf{X}$ is defined as follows:
\begin{align}
	\mathbf{X} = \begin{bmatrix}
	\mathbf{x}[k] &  \mathbf{x}[k + 1] & \ldots & \mathbf{x}[N]
	\end{bmatrix}^T \label{eq:nf_stack}
	\end{align}
The problem of estimating parameter vector ($\mathbf{\theta}$) from \emph{noise-free} measurements can be solved using multiple frameworks such as ordinary least squares \citep{paper:xu1995least} and PCA which minimizes the total least squares (TLS) error function \citep{book:joliffe}. To illustrate the above idea consider a second order LTI system as shown below: 
 \begin{align}
	y^{\star}[k] &- 0.4y^{\star}[k-1] + 0.6y^{\star}[k-2] = 2u^{\star}[k-1] \label{eq:secondord_sys}
	\end{align}
The input $u^{\star}$ is chosen to be a full band pseudo random binary signal (PRBS) of length $1023$. The output $y^{\star}$ is generated according to Eqn. \eqref{eq:secondord_sys}. The next step is to apply PCA by using eigenvalue decomposition (EVD) on the sample covariance matrix, $\mathbf{S_X} = \frac{1}{N-\eta} 	\mathbf{X}^T 	\mathbf{X}$ as shown below:
\begin{align}
\mathbf{S_X} \mathbf{V^{\star}_0} &= \mathbf{V^{\star}_0}  \mathbf{\Lambda^{\star}_0 } \label{eq:eig_nf}
\end{align}
where $\mathbf{\Lambda^{\star}_0 }$, $\mathbf{V^{\star}_0}$ consists of eigenvalues and corresponding eigenvectors respectively. By performing EVD, the minimum eigenvalue is observed to be $0$ and the corresponding eigenvector is 
 \begin{align}
	\mathbf{V^{\star}_{0}} &= \begin{bmatrix}
	0.4256   & -0.1703  &  0.2554 &  -0.8513
	\end{bmatrix} \label{eq:noisefree_eigvec}	
	\end{align} 
The estimate of parameter vector can be obtained from above by normalizing the coefficient of dependent variable $y[k]$ to be unity: 
\begin{align}
	\mathbf{\theta} = \begin{bmatrix}
	1 &   -0.4 &  0.6 &  -2
	\end{bmatrix}^T
	\end{align}
It can be observed that the estimated parameters are same as used in simulating the data in Eqn. \eqref{eq:secondord_sys}. 
%One of the key assumptions is that the orders ($n_a, n_b$) and delay ($D$) are known a priori which is usually not the case. Dynamic PCA \cite{paper:dpca} attempted the above problem assuming the orders are unknown. In such a scenario, both the input and output lagged variables are stacked in equal orders. The key idea was to sequentially try every guess of maximum order, $\eta$ until the recognition of at least one constraint. The number of constraints can be guessed from the analysis of \emph{zero} singular values for \emph{noise-free} measurements. 

The entire identification task becomes challenging for noisy output measurements.  PCA and its variants deliver unbiased estimates of parameter vector-only for \emph{homoskedastic} errors case, meaning both output ($y^{\star}$) and input ($u^{\star}$) are contaminated by Gaussian white noise of same variance \citep{book:joliffe}. 

In our case, the output noise is colored, and the input is noise-free. In the next subsection, we discuss more generalized variants of PCA which can handle some of these issues. 

\subsection{Iterative PCA \& Dynamic IPCA}
\label{sec:ipca_dipca}
In this subsection, we discuss the key idea to handle \emph{heteroskedastic} errors used in \cite{paper:iecr_dipca} and make relaxing assumptions that the order and noise variances are known. It was originally proposed for different settings but we discuss only the few relevant aspects for the proposed work. 

Iterative PCA \cite{paper:ipca} was proposed to handle the \emph{heteroskedastic} errors case. Dynamic iterative PCA referred as DIPCA \citep{paper:iecr_dipca} was proposed recently and can be viewed as nature extension of IPCA for dynamic models. One of the novel contributions of these work is order determination by analysis of singular values. 

\begin{figure}[htb]
	\begin{minipage}[b]{1.0\linewidth}
		\centering
		\centerline{\includegraphics[width=7cm]{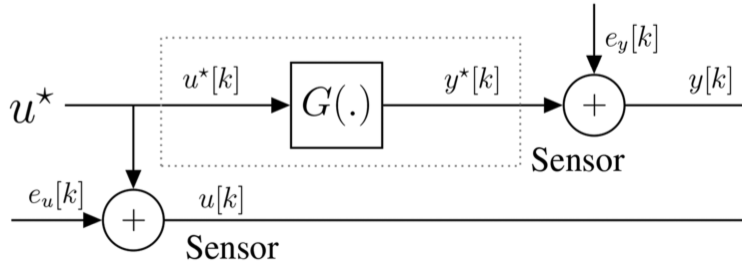}}
	\end{minipage}
	\caption{EIV Model Architecture}
	\label{fig:eiv_process}
\end{figure}

Dynamic IPCA \citep{paper:iecr_dipca} algorithm assumed the presence of noise in both input and output variables as shown in Figure \ref{fig:eiv_process}. This is contrary to our settings as the input is assumed to be noise-free. But these are discussed briefly to understand the prolific merits of handling \emph{heteroskedasticity} and order determination using the EVD framework, as shown later in Section \ref{sec:prop_2}. 

Consider both the input and output measurements are corrupted by Gaussian white noise of different variances as shown below:
 \begin{align*}
	y[k] &= y^{\star}[k] + e_y[k] \\ 
	u[k] &= u^{\star}[k] + e_u[k]
	\label{eq:eiv_assump}
	\end{align*}
where $e_y[k] \sim \mathcal{N}(0,\sigma^2_{e_y})$, $e_u[k] \sim \mathcal{N}(0,\sigma^2_{e_u})$. The collection of $N$ measurements for this case can be given by: 
{\scriptsize \begin{subequations}
		\begin{align}
		\mathbf{z}[k]  &= \begin{bmatrix}
		y[k]  & \ldots & y[k - n_y] & u[k-D]  & \ldots& u[k-n_u]
		\end{bmatrix}^T \\
		\mathbf{Z} & = \begin{bmatrix}
		\mathbf{z}^T[\eta] &  \mathbf{z}^T[\eta + 1] & \ldots & \mathbf{z}^T[N]
		\end{bmatrix}
		\end{align} 
		\label{eq:lagconstr_known}
\end{subequations}}
The noise covariance is a diagonal matrix of dimension $(n_a + n_b - D + 1) \times (n_a + n_b - D + 1)$ with its entries as, $\text{diag}(\mathbf{\Sigma_e}) = \begin{bmatrix}
\sigma^2_{e_y} \mathbf{1}_{n_a} & \sigma^2_{e_u} \mathbf{1}_{n_b - D + 1}
\end{bmatrix}$, where $\mathbf{1}_m$ denotes a vector containing all the $m$ elements as unity. 

The key idea in \cite{paper:ipca,paper:iecr_dipca} is to map the data from \emph{heteroskedastic} errors to \emph{homoskedastic} space by scaling the data with inverse square root of noise covariance matrix which is assumed to be known. \citet{paper:iecr_dipca} describes the method to estimate noise covariance matrix also but we skip that discussion as it is felt to be inessential for this work. 
\begin{align}
	\mathbf{L} 	\mathbf{L}^T &=  \mathbf{\Sigma_e} \nonumber \\  
	\mathbf{Z_s} &= 	\mathbf{Z} \mathbf{L}^{-1} 	
	\label{eq:scale_dipca}
	\end{align}
The next step is to perform spectral decomposition on the sample covariance matrix of the scaled data matrix, $\mathbf{S_{Z_s}} = \frac{1}{N-\eta}\mathbf{Z_s}^T\mathbf{Z_s}  $ as shown below:
\begin{align}
	\mathbf{S_{Z_s} V_s} &= \mathbf{V_s \Lambda_s} \label{eq:scale_eig}
	\end{align}
where $\mathbf{V_s}$ and $\mathbf{\Lambda_s}$ denote the eigenvectors and eigenvalues in scaled domain. It is theoretically proved in \cite{paper:ipca} that the eigenvectors in scaled domain can be mapped to original space using 
\begin{align}
	\mathbf{\Lambda_s} &= \mathbf{\Lambda^{\star}_0 } + \mathbf{I} \\ 
	\mathbf{V_s} &=  \mathbf{L}^{-1} \mathbf{V^{\star}_0} 
	\end{align}
where $\mathbf{I} $ denotes identity matrix. The above equations state the zero eigenvalues of $\mathbf{S_x}$ map to unity in the scaled domain, which helps in order determination. The parameter vector can also be recovered from the eigenvector in the scaled domain. 

The above discussion was carried under the assumption that both input-output variables contain Gaussian white noise, but in our case, only one of the variables - output has colored noise and input is noise-free. In the next section, we describe the proposed method which handles this issue. 

\section{Proposed Method}
\label{sec:prop}
From the discussion in the previous section, we have concluded that three main challenges for our case are
\begin{enumerate}
    \item Handling of colored noise in \textit{output only}.
    \item Order and delay determination 
    \item Model parameter estimation 
\end{enumerate}
To address these challenges one by one, we present the proposed algorithm in three subsections. 
\subsection{Model identification for known order \& ACVF of Noise}
\label{sec:prop_1}
In this subsection, we first draw the equivalence of scaling approach in \cite{paper:iecr_dipca} to the generalized spectral decomposition. By using some elementary linear algebra properties, it can be shown that Eqn. \eqref{eq:scale_dipca} and Eqn. \eqref{eq:scale_eig} can be expressed as:
\begin{align}
\mathbf{S_{Z} V} &= \mathbf{\Sigma_e}\mathbf{V \Lambda_s} \label{eq:gen_eig_1}
\end{align}
In this subsection as the ACVF of noise is assumed to be known, we construct appropriate form of $\mathbf{\Sigma_e}$ and utilize generalized spectral decomposition in Eqn. \eqref{eq:gen_eig_1}. As seen in Eqn. \eqref{eq:arx_noise}, the noise $v[k]$ is colored. 
%So by noise distribution, we refer to the knowledge of auto-covariance function (ACVF) of $v[k]$. 

To illustrate the key idea reconsider the same system in Eqn. \eqref{eq:secondord_sys}. Noisy output measurements ($y$ in Eqn. \eqref{eq:yk_noise}) are generated in accordance to  Eqn. \eqref{eq:secondord_sys} and Eqn. \eqref{eq:arx_noise} with noise variance of $\sigma_{e}^2 = 0.4 $. The corresponding value of noise variance was chosen to maintain $\frac{\text{var}(y^{\star})}{\text{var(v)}} = 10$. For stacking of lagged input-output variables shown in Eqn. \eqref{eq:lagconstr_known}, the noise covariance is shown below:  
\begin{align}
	\mathbf{\Sigma_e} = \begin{bmatrix}
	\sigma_{vv}[0] &\sigma_{vv}[1]  &\sigma_{vv}[2] & 0 \\
	\sigma_{vv}[1] &\sigma_{vv}[0]  &\sigma_{vv}[1] & 0 \\
	\sigma_{vv}[2] &\sigma_{vv}[1]  &\sigma_{vv}[0] & 0 \\
	0 &            0              & 0  & 0 \\
	\end{bmatrix}
	\label{eq:sigma_exmp}
	\end{align}
where $\sigma_{vv}[l]$ denotes ACVF for $v[k]$at lag $l$. The next step is to preform spectral decomposition of the sample covariance matrix. Equation Eqn. \eqref{eq:gen_eig_1} reduces to usual eigenvalue problem if the inverse of noise covariance matrix exists. But under the assumption of input being noise-free,  $\mathbf{\Sigma_e}$ constructed in Eqn. \eqref{eq:sigma_exmp} will always be singular. 

In order to solve Eqn. \eqref{eq:sigma_exmp}, we utilize the QZ algorithm \cite{paper:qz_moler} which is briefly outlined in Appendix. It can be observed as a powerful tool that utilizes simple row operations to solve the generalized EVD problem even for singular $\mathbf{\Sigma_e}$. It is a generalization of a well-known QR algorithm and we encourage the reader to refer \cite{paper:qz_moler} for more details. 

On applying the QZ algorithm for our example, the minimum generalized eigenvalue is observed to be $0.98$, and the parameter vector computed from corresponding generalized eigenvector is:
 \begin{align}
	\mathbf{\hat{\theta}} = \begin{bmatrix}
	1 & -0.4059  & 0.6036 & -2.018
	\end{bmatrix}
	\end{align}
It can be observed that the estimated parameters are close to the true values used in Eqn. \eqref{eq:secondord_sys} and the minimum eigenvalue is unity. This establishes the estimates of desired parameters can be obtained for known order and noise characteristics.  

\subsection{Estimation of order for known ACVF of noise}
\label{sec:prop_2}
The discussion in previous subsection was conditioned on the  availability of order and noise characteristics. In this subsection, we discuss the key idea for estimation of order for given ACVF of noise. 

When the order is unknown, the input-output variables should be stacked up to lag order $L$ as shown below: 
\begin{subequations}
		\begin{align}
		z_{L}[k] & = \begin{bmatrix} y[k] &  y[k-1] \cdots & y[k-L]   & u^{\star}[k] \cdots & u^{\star}[k-L]
		\end{bmatrix}^T \label{eq:zk_def_dyn_gen_L}  \\
		\mathbf{Z}_L & = \begin{bmatrix} z_{L}[L] & z_{L}[L+1] & \cdots & z_{L}[N]
		\end{bmatrix}^T  \label{eq:Z_data_gen_L} 
		\end{align}
\end{subequations} 
The number of identified linear relations ($\hat{d}$) for chosen stacking order ($L$) could potentially lead to following three cases:
\begin{enumerate}
	\item $L < \eta $: If the maximum lag of stacked variables is less than equation order, then there would be no linear relations and hence $\hat{d} = 0$. This is the under-parameterization case which could be detected from the absence of linear relations among stacked variables. 
	\item $L = \eta $: As discussed in the previous sub-section, there would be only one linear relation. This is the ideally desired choice of stacking order but we have assumed the equation order to be unavailable and hence can not be used as stacking order. 
	\item $L > \eta $: If the input-output variables are stacked in excess, multiple linear relations would be detected. This implies $d > 1$.  
\end{enumerate} 
The stacking order, $L$ is chosen to be at least equation order ($\eta$). This is done to identify at least one linear relation among the columns of $	\mathbf{Z}_L$. The number of excessively identified relations could be potentially used to estimate the equation order, which is described in this subsection.

The  sample covariance matrix for $\mathbf{Z}_L$ is denoted by $\mathbf{S_{Z_L}}$ and the full noise covariance matrix is given by:
	\begin{align}
\mathbf{S_{Z_L}} &=  \frac{1}{N-L}\mathbf{Z_L}^T\mathbf{Z_L} \\
\mathbf{\Sigma_{e_L}} &= \begin{bmatrix}
\mathbf{ \Sigma_{v}}	& \mathbf{0}_{P} \\ 
	\mathbf{0}_{P} & \mathbf{0}_{P}
	\end{bmatrix}
	\label{eq:noise_covarx}
	\end{align}
where $\mathbf{0}_{P}$ denotes a zero matrix of dimension $(L+1) \times (L+1)$ and $\mathbf{ \Sigma_{v}}$  is a the output noise covariance matrix. 
\begin{align}
\mathbf{ \Sigma_{v}}=	\begin{bmatrix}
	\sigma_{vv}[0] & \sigma_{vv}[1] & \sigma_{vv}[2]  & \ldots &\sigma_{vv}[L] \\
	\sigma_{vv}[1] & \sigma_{vv}[0] & \sigma_{vv}[1]  & \ddots & \sigma_{vv}[L-1] \\
	\sigma_{vv}[2] & \sigma_{vv}[1] & \sigma_{vv}[0] & \ddots & \vdots \\
	%\sigma_{vv}[3] & \sigma_{vv}[2] & \ddots & \ddots & \ddots & \vdots \\
	\vdots &  \ddots & \ddots &  \ddots  & \vdots  \\
	\sigma_{vv}[L] & \sigma_{vv}[L-1] & \ldots & \ldots & \sigma_{vv}[0]
	\end{bmatrix} 
\end{align}
It can be easily noticed $\mathbf{ \Sigma_{v}}$ is a symmetric Toeplitz matrix. 

The next step is to utilize the generalized eigenvalue decomposition for identification of linear relations:
 \begin{align}
\mathbf{S_{Z_L} V} &= \mathbf{\Sigma_{e_L}} \mathbf{V \Lambda_s} \label{eq:gen_eig_2}
\end{align}
We reconsider the same case study described in Eqn. \eqref{eq:secondord_sys} for $L = 3$. As the stacking order for lagged variables is greater than equation order, we should expect excess linear relations, which are as follows:
\begin{enumerate}
	\item the first linear relation among the variables $y[k]$, $y[k-1]$, $y[k-2]$ and $u[k-1]$. 
	\item the second linear relation among the delayed version of the same variables, which are $y[k-1]$, $y[k-2]$, $y[k-3]$ and $u[k-2]$. 
\end{enumerate}
It should be noticed that the identified linear relations could also be linear combination of the above two constraints. This is due to the fact that generalized EVD provides a basis for the constraint matrix. 

In Section \ref{sec:ipca_dipca}, it was discussed that the zero eigenvalues map to unity  in scaled domain which helps in order determination.

The last two eigenvalues are observed to be in close proximity of unity, which provides clue for the existence of two linear relations among the lagged input-output variables. For a SISO system, as there exists only one unique linear relationship, the equation order can be estimated from
 \begin{align}
	\hat{\eta} = L - \hat{d} + 1 \label{eq:ord_det}
	\end{align}
where $\hat{d} $ denotes the number of unity eigenvalues. Using this the order is estimated to be $\hat{\eta} = 3 - 2 + 1 = 2$. The key idea of order determination is inspired from \cite{paper:ipca,paper:iecr_dipca}.   
\subsection{Estimation of order \& ACVF of noise}
\label{sec:prop_3}
In this subsection, the discussion is constructed on realistic assumption of non-availability of any prior information about order and ACVF of noise. The key idea is to use the discussion in Sections \ref{sec:prop_1} and \ref{sec:prop_2} iteratively. One important aspect for this iterative algorithm is estimation of ACVF of noise for given order. Equation \eqref{eq:dgp} can be interpreted as 
\begin{align}
	\begin{bmatrix}
	A(z^{-1}) & - B(z^{-1})
	\end{bmatrix} 
	\begin{bmatrix}
	y[k] \\
	u^{\star}[k]
	\end{bmatrix} &= e[k] \label{eq:res}
	\end{align}
For a chosen guess of equation order $\eta$, we could obtain the estimates of $\begin{bmatrix}
A(z^{-1}) & - B(z^{-1})
\end{bmatrix} $ using the last generalized eigenvector as described in Section \ref{sec:prop_1}. Equation Eqn. \eqref{eq:res} could also be used to obtain the noise variance estimate $\sigma_e^2$. 

The next step is to obtain the estimate  ACVF of $v[k]$ for construction of $ \mathbf{\Sigma_{e}}$ in Eqn. \eqref{eq:gen_eig_1}. This can be done using Wiener-Khinchin theorem \citep{book:akt_sysid,book:ljung}. Any stationary process with ACVF $\sigma_{vv}[l]$ satisfying $\sum_{l=-\infty}^{\infty} |\sigma_{vv}[l]| < \infty $ has the following spectral representation 
\begin{subequations}
		\begin{align}
		\sigma_{vv}[l] &= 	\int_{-\pi}^{\pi} \gamma_{vv}(\omega)e^{j\omega l} d\omega \\
		\gamma_{vv}(\omega) &= \frac{1}{|A(e^{-j\omega})|^2}\gamma_{ee}(\omega) = \frac{1}{|A(e^{-j\omega})|^2}\frac{\sigma^2_{e}}{2\pi} 
		\end{align}
		\label{eq:weiner}
\end{subequations}
This can be used to estimate the ACVF function of $v[k]$. The next step is to re-apply spectral decomposition for assumed equation order as shown in Eqn. \eqref{eq:gen_eig_1} to obtain a refined estimate of parameter vector. Using this key idea, proposed  algorithm is summarized in Table \ref{tab:qz_arx}.

\begin{table}[thpb]
	\caption{ARX model identification using generalized spectral decomposition}
	\label{tab:qz_arx}
	\vskip -3mm
	\hrulefill
{\normalsize \begin{enumerate}
		\itemsep0.5em 
		\item The first step is to construct the data matrix as shown in Eqn. \eqref{eq:Z_data_gen_L}. 
		\item For iteration $i = 0$, kick start the algorithm with some guess of equation order ($\hat{\eta}$) and noise covariance matrix ($\mathbf{\Sigma_{e}}$). An easy approach is to use $\mathbf{\Sigma_{e}} = \mathbf{I}$  and apply generalized EVD using Eqn. \eqref{eq:gen_eig_1}. 
		\item Estimate the ACVF as discussed in Eqn. \eqref{eq:weiner} using Wiener-Khinchin theorem. Re-perform generalized EVD to obtain refined estimate of parameter vector using updated $\mathbf{\Sigma_{e}}$ in Eqn. \eqref{eq:gen_eig_1}. Repeat step 2 for few iterations until the convergence of parameter vector or ACVF of $v[k]$.
		\item For the converged estimate of ACVF of $v[k]$, reconstruct $\mathbf{Z}_L$ as shown in Eqn. \eqref{eq:Z_data_gen_L} and re-apply generalized EVD using Eqn. \eqref{eq:gen_eig_2} to obtain an estimate for the number of linear relations ($d$).
		\item Check if Eqn. \eqref{eq:ord_det} holds true. If yes, the overall algorithm has converged and if not, repeat the entire procedure from step 1 with modified guess of $\eta$.
	\end{enumerate}}
	\hrulefill
\end{table}

In this section, we discussed the proposed algorithm for the identification of all the parameters of an ARX model under the assumption of the non-availability of any prior information. All the parameters are derived systematically in the proposed algorithm. The prime contribution of this work is the estimation of the equation order. The proposed algorithm utilizes the generalized spectral decomposition framework, which encapsulates the order determination and model parameter estimation in a single step. As a byproduct, ACVF of noise can also be estimated. This step of order determination plays a crucial role as it avoids the need for checking the over-parameterization.
% Theoretical proof for order determination is presented in \cite{paper:iecr_dipca} for error-in-variables case and is under construction for our case. provides .. avoid over fitting check

The simulation results are presented in the next section to demonstrate the functioning of proposed algorithm.

\section{Simulation Results}
\label{sec:simulation}
\subsection{Second-Order LTI system}
\begin{figure}
\begin{center}
\includegraphics[width=8.4cm]{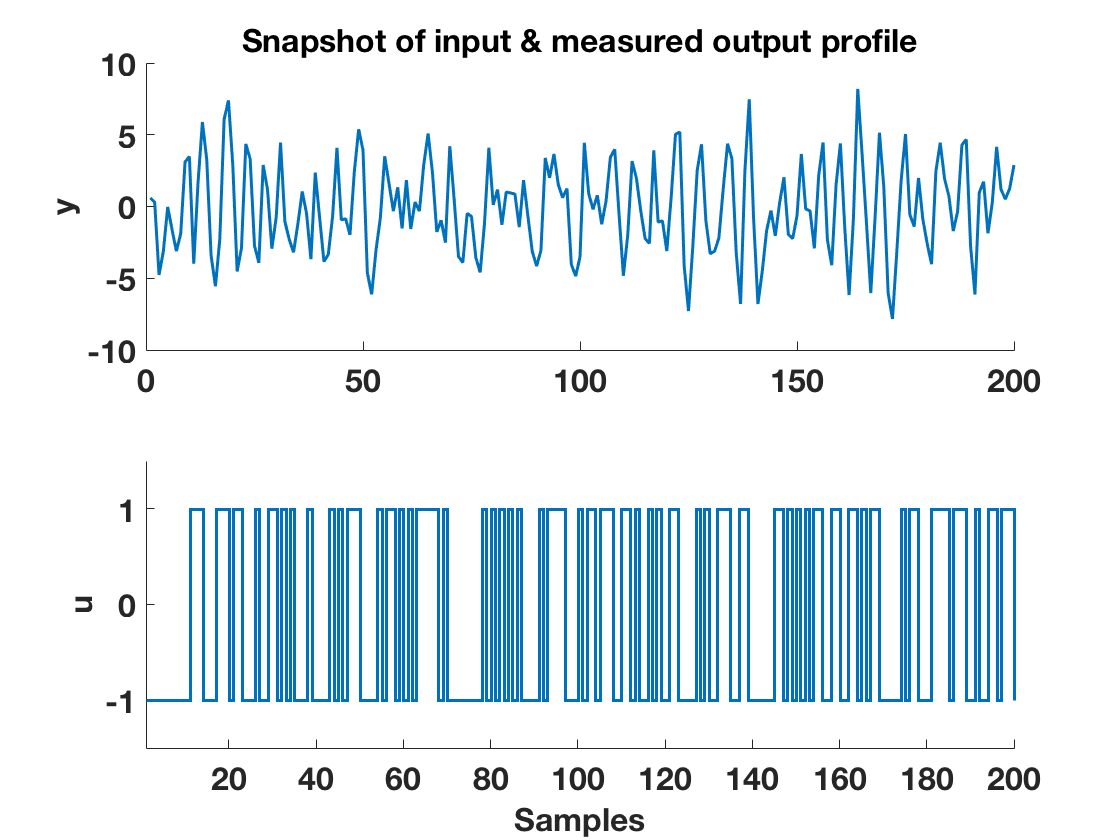}    % The printed column width is 8.4 cm.
\caption{Input-Output Data}
	\label{fig:io_2nd}
\end{center}
\end{figure}

We reconsider the same second order system described in Eqn. \eqref{eq:secondord_sys}. The data generating process is same as mentioned in Section \ref{sec:prop_1} and a snapshot of input-output data is shown in Figure \ref{fig:io_2nd}. 

\begin{figure}
\begin{center}
\includegraphics[width=8.4cm]{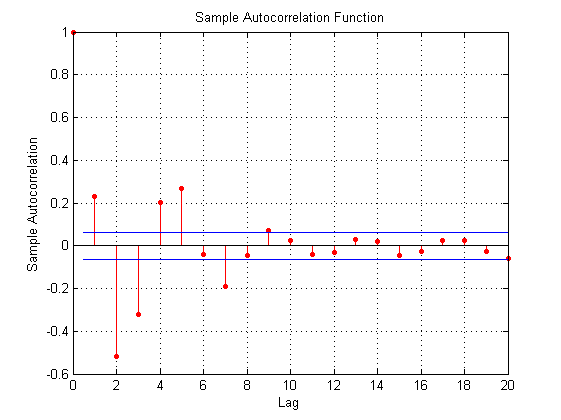}    % The printed column width is 8.4 cm.
\caption{ACVF of output noise}
	\label{fig:acf_noise}
\end{center}
\end{figure}

The auto-correlation function ACF of output noise is shown in Figure \ref{fig:acf_noise}. It demonstrates the noise is colored. The simulated noise variance was chosen as $1.4368$ to maintain SNR = 5. 

% Through this example, we demonstrate the functioning of proposed algorithm in various scenarios. We start the estimation process under the assumption with no prior information available. 
As the order is assumed to be unknown, we consider the case with $\eta_{\text{guess}} = 1$. So the data matrix is constructed with $L = 1$ and first three steps of proposed algorithm are applied.
% As we have assumed no information about the system properties are unavailable a priori, we kick start the algorithm with guess of equation order as $\eta = 1$. 
The parameter estimates and the last 3 eigenvalues obtained at the convergence:
 \begin{subequations}
		\begin{align}
		\mathbf{\hat{\theta}} &= \begin{bmatrix}
		1 & -0.2242  & -0.0012 &-1.96
		\end{bmatrix} \\
		\mathbf{\Lambda_s} &= \begin{bmatrix}
		&  0.3448  &  0.2026   & 0.1546
		\end{bmatrix}
		\end{align}
\end{subequations}
As none of the eigenvalues are unity, we discard $\eta_{\text{guess}} = 1$.

The next step is to increase the guessed order and choose $\eta_{\text{guess}} = 2$. The estimated parameter and last $5$ eigenvalues for $L = 5$ after convergence:
\begin{subequations}
		\begin{align}
		\mathbf{\hat{\theta}}&= \begin{bmatrix}
		1 &  -0.409 &   0.611  & -0.004 &-1.969 &   0.007
		\end{bmatrix} \\
		\mathbf{\Lambda_s} &= \begin{bmatrix}
		4.5536  &  1.0688 &   1.0493 &   0.9988  &  0.9689
		\end{bmatrix}
		\end{align}
\end{subequations}
As there are $4$ unity eigenvalues, the equation order is estimated to be $\hat{\eta} = 5-4+1 = 2$. It can also be observed that the estimated parameter vector is in close agreement with Eqn. \eqref{eq:secondord_sys}. Bootstrap simulations \citep{shumway2000time} are performed to derive the confidence interval of model parameters:
\begin{align} 
	y[k] - \underset{\pm (0.023 )}{0.399 }y[k-1] + \underset{ \pm (0.021)}{ 0.601}y[k-2] =  
	\underset{\pm (0.036)}{2.003}u[k-1]  \label{eq:final_est}
	\end{align}
The noise variance also estimated to be $1.459$, which is close to the simulated value. The ACVF of noise is also estimated accurately as the parameter estimates of difference equation are close to simulated theoretical values. 

As the $\hat{\eta}  = \eta_{\text{guess}} = 2$, the algorithm is converged, but for the sake of completeness, we consider the case of  $\eta_{\text{guess}} = 3$. The last $3$ generalized eigenvalues for $L = 5$ are shown below:
\begin{align}
\mathbf{\Lambda_s} =  \begin{bmatrix} 1.09   & 0.29   &  0.0068 \end{bmatrix}
\end{align}
As the number of unity eigenvalues is not $3$, we discard $\eta_{\text{guess}} = 3$. The above exercise demonstrates that a wrong guess of order can be detected by inspection of generalized eigenvalues. It should be noticed that the proposed framework provides the clues for over-parameterization and under-parameterization, which helps converging to the right order quickly. 

In order to demonstrate the performance of proposed algorithm at higher levels of noise, we consider SNR = 3. So the output noise variance is chosen accordingly. 
We repeat the entire procedure of estimating the model parameters with stacking order of $L = 5$ and $\eta_{guess} = 2$. We obtain $4$ unity eigenvalues and the estimated order is found to be $\hat{\eta} = 5 - 4 + 1 = 2$. The final estimated difference equation is shown below: 
\begin{align*} 
	y[k] - \underset{\pm (0.096 )}{0.403}y[k-1] + \underset{ \pm (0.064)}{ 0.596}y[k-2] =  
	\underset{\pm (0.068)}{2.007}u[k-1] 
	\end{align*}
It should be noticed that the variance in estimated parameters is higher for this case compared to  Eqn. \eqref{eq:final_est} due to low SNR. This shows the appreciable quality model parameter estimates can be obtained even at low SNR. 

In order to compare the proposed algorithm to existing methods, we estimate the parameters using ordinary least squares (OLS) for this ARX model. The estimates are reported as 
\begin{align*} 
y[k] - \underset{\pm (0.012 )}{0.414}y[k-1] + \underset{ \pm (0.0121)}{ 0.611}y[k-2] =  
\underset{\pm (0.035)}{2.002}u[k-1] 
\end{align*}

It can be observed that the quality of estimated parameters is equivalent from both the methods. It should be noticed that the proposed framework is capable of deriving the equation order also where the user is forced to supply the right combination of input-output order and delay to OLS algorithm. 
% TODO: summarize the contributions , put another example  

\subsection{ Third Order LTI System}
Consider a second order system with third order input dynamics shown below:

{\footnotesize \begin{align}
		y^{\star}[k] &- 0.3y^{\star}[k-1] + 0.7y^{\star}[k-2] = 1.2u^{\star}[k-2]  + 1.6u^{\star}[k-3] \label{eq:3rd_sys}
\end{align}}
Through this example, we show the functioning of proposed algorithm for the cases where input order is greater than output order. 

\begin{figure}
\begin{center}
\includegraphics[width=8.4cm]{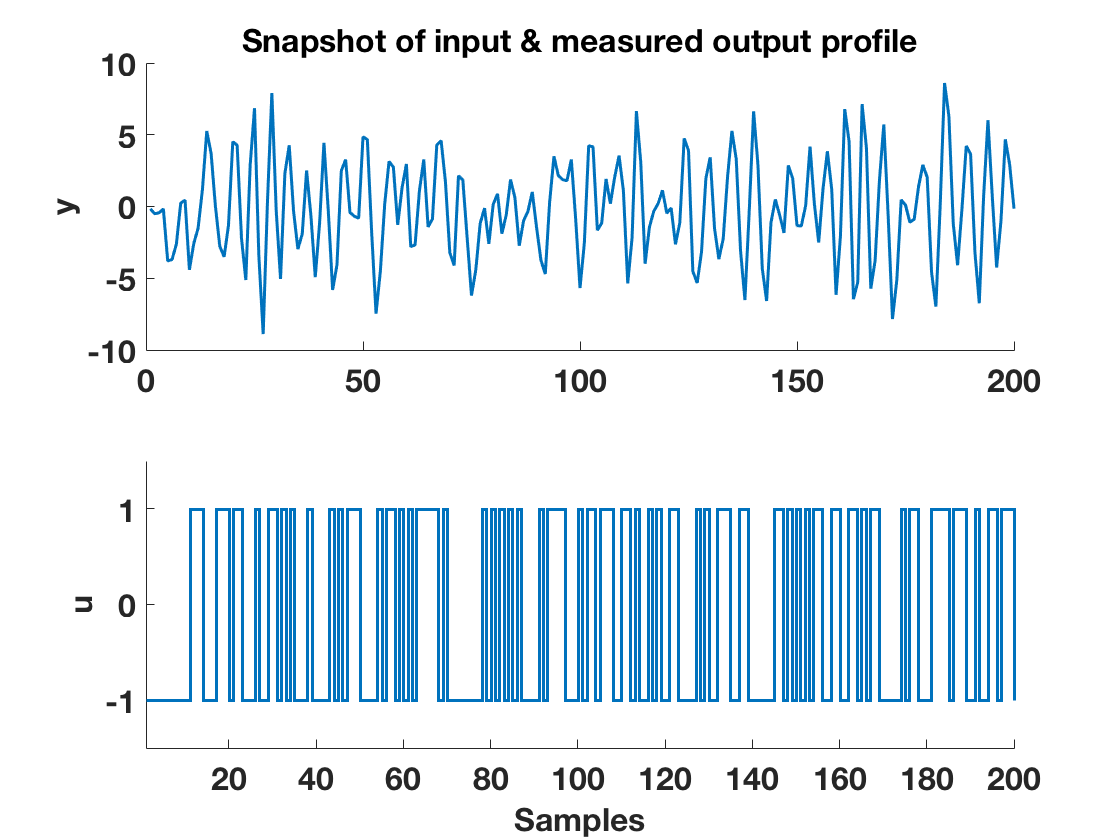}    % The printed column width is 8.4 cm.
\caption{Input-Output Data}
	\label{fig:io_3rd}
\end{center}
\end{figure}

Input is chosen to be full length and full band PRBS signal with sample size as 1023. The noise free output is generated according to Eqn. \eqref{eq:3rd_sys} and further it is corrupted with noise to follow ARX model. The noise variance is chosen to be $1.7$ to maintain SNR of $6$. A snapshot of the input-output data is shown in Figure \ref{fig:io_3rd}.

\begin{figure}[htb]
	\begin{minipage}[b]{1.0\linewidth}
		\centering
		\centerline{\includegraphics[width=6.3cm]{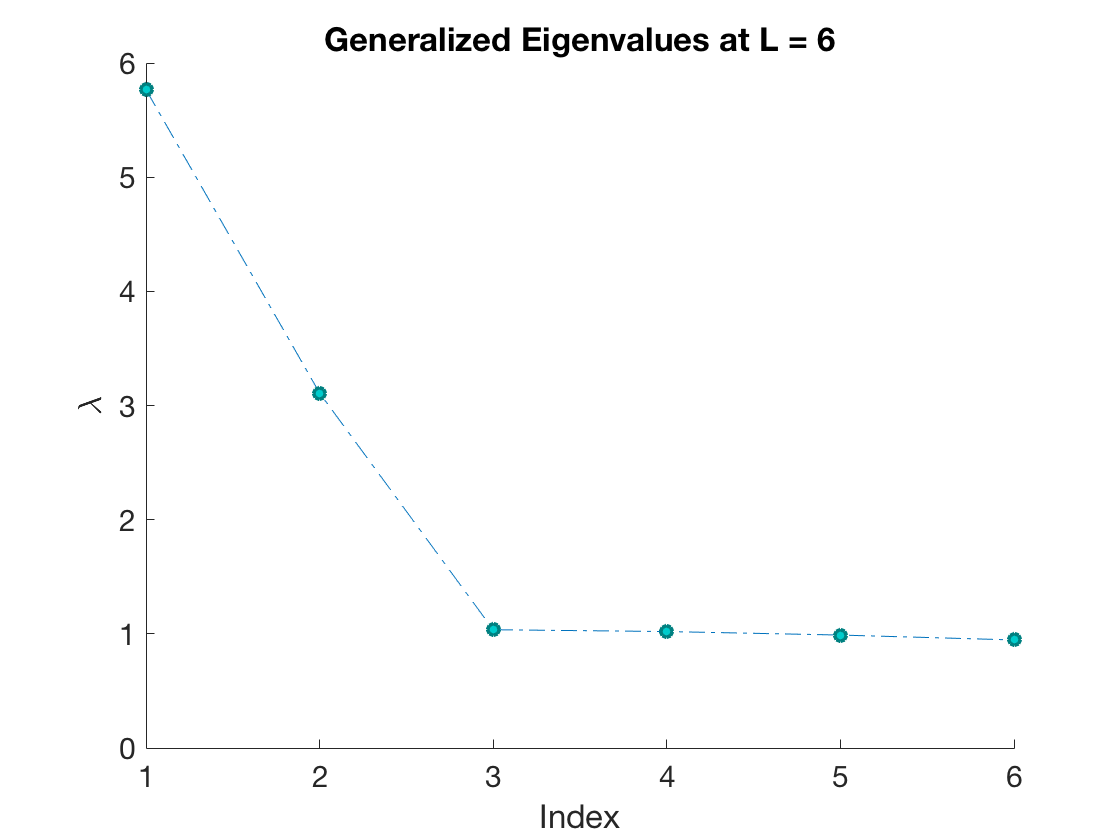}}
		%  \vspace{2.0cm}
	\end{minipage}
	\caption{Converged Eigenvalues}
	\label{fig:eig_cs2}
\end{figure}

As the order is assumed to be unknown, we consider high stacking lag order of $L = 6$, which implies there would be $2(L + 1) = 14$ variables. The guessed order is chosen as $ \eta_{\text{guess}} = 3$. The last $6$ converged eigenvalues  are shown in Figure \ref{fig:eig_cs2}.

It can be observed that the last $4$ eigenvalues are unity. So the estimated order would be $\hat{\eta} = L - \hat{d} + 1 = 6 - 4 + 1 = 3$, which is also equal to the maximum lag in the difference equation shown in Eqn. \eqref{eq:3rd_sys}. The estimated noise variance is $1.75$ which is also close to the true simulated value.

The convergence of estimated parameters with each iteration is shown in Table \ref{tab:conv_2}. 
 
\begin{small}
	\begin{table}[thpb]
		\caption{Convergence of Estimates \label{tab:conv_2}}
		\vspace{-0.25cm}
		\centering
		\begin{tabular}{*2c}
			\hline
			Iter. No.            & $\mathbf{ \hat{\theta}} = \begin{bmatrix}
			1 & \hat{a}_1 & \hat{a}_2 & \hat{a}_3 & -\hat{b}_0 &  -\hat{b}_1 &  -\hat{b}_2 &  -\hat{b}_3  
			\end{bmatrix}$     \\ \hline
			1 & $\begin{bmatrix}
			1 &   -0.25 &    0.65 &    0.05 &   0.03 &  0.01 &  1.19 &   1.67 
			\end{bmatrix}$           \\ 
			2 & $\begin{bmatrix}
				1 &   -0.36 &    0.73 &   -0.04 &  0.03 &  0.01 &  1.2  & 1.55
			\end{bmatrix}    $       \\
			3 & $ \begin{bmatrix}
				1 &   -0.30 &   0.70 &    0.01 & 0.03 &   0.01 &   1.2 &   1.61  
			\end{bmatrix}     $     \\
			4 & $ \begin{bmatrix}
			1 &   -0.33   & 0.70 & -0.02 &  0.03 &  0.01 &  1.19  & 1.58
			\end{bmatrix}     $     \\
			5 & $ \begin{bmatrix}
			1 &   -0.31 & 0.70 &   -0.01 & 0.03 &   0 &  1.19 &   1.59
			\end{bmatrix}     $     \\
			\hline
		$\mathbf{ \theta}$ & $\begin{bmatrix}
				1 &   -0.3 & 0.7 &   0 & 0 &   0 &  1.2 &   1.6
			\end{bmatrix}     $ \\
			\hline
		\end{tabular}	
	\end{table}
\end{small} 
As the estimated order is observed to be equal to guessed order, we terminate the algorithm. It can be clearly observed that the estimated parameters is close to the true simulated values used in simulation from Table \ref{tab:conv_2}. The last row in Table \ref{tab:conv_2} contains the true parameters of the difference equation specified in Eqn. \eqref{eq:3rd_sys}. 

Using the estimated model parameters in Eqn. \eqref{eq:final_est}, we evaluate the percentage fit from noise-free output and is observed to be $91\%$. 
%This above exercise shows the adaptive noise cancellation could be achieved to an appreciable quality by using the proposed algorithm which derives the model orders, delay and filter parameters in an automated manner. 
The above simulations show that model order and parameters can be derived by careful inspection of generalized eigenvalues and eigenvectors.  
%eigenvalues and parameter vector: not satisfactory 
%for eta = 2, parameter and eigenvalues... stack and reconfirm for L = 5. Could stop here
% for discussion eta = 3, show eigenvalues hint for 2 
In this section, we considered two case-studies and demonstrated working of each step of the proposed algorithm. The notable takeaways are estimation of equation order  and efficacy of the obtained estimates in automated manner without supplying any prior knowledge about the system.

\section{Conclusion} 
\label{sec:conc}
In this article, we proposed a novel automated framework which derives the equation order and model parameters using generalized spectral decomposition framework. The equation order is determined by analyzing the generalized eigenvalues and subsequently the model parameters of linear difference equation can be obtained from generalized eigenvector. The algorithm can be viewed as a two step iterative procedure. 
%The proposed algorithm can be viewed as generic framework which caters whose application is not limited to ANC. 
Future work consists of extending the proposed framework to time-varying systems \cite{soton425875} and from single channel to multichannel with more generalized and complex  model structure Box-Jenkins model. 

%\vfill\pagebreak

\bibliography{myref}             % bib file to produce the bibliography

\begin{thebibliography}{4}
\providecommand{\natexlab}[1]{#1}
\providecommand{\url}[1]{\texttt{#1}}
\providecommand{\urlprefix}{URL }
\expandafter\ifx\csname urlstyle\endcsname\relax
  \providecommand{\doi}[1]{doi:\discretionary{}{}{}#1}\else
  \providecommand{\doi}{doi:\discretionary{}{}{}\begingroup
  \urlstyle{rm}\Url}\fi

\bibitem[{Able(1956)}]{Abl:56}
Able, B. (1956).
\newblock Nucleic acid content of microscope.
\newblock \emph{Nature}, 135, 7--9.

\bibitem[{Able et~al.(1954)Able, Tagg, and Rush}]{AbTaRu:54}
Able, B., Tagg, R., and Rush, M. (1954).
\newblock Enzyme-catalyzed cellular transanimations.
\newblock In A.~Round (ed.), \emph{Advances in Enzymology}, volume~2, 125--247.
  Academic Press, New York, 3rd edition.

\bibitem[{Keohane(1958)}]{Keo:58}
Keohane, R. (1958).
\newblock \emph{Power and Interdependence: World Politics in Transitions}.
\newblock Little, Brown \& Co., Boston.

\bibitem[{Powers(1985)}]{Pow:85}
Powers, T. (1985).
\newblock Is there a way out?
\newblock \emph{Harpers}, 35--47.

\end{thebibliography}

\begin{thebibliography}{27}
\providecommand{\natexlab}[1]{#1}
\providecommand{\url}[1]{\texttt{#1}}
\providecommand{\urlprefix}{URL }
\expandafter\ifx\csname urlstyle\endcsname\relax
  \providecommand{\doi}[1]{doi:\discretionary{}{}{}#1}\else
  \providecommand{\doi}{doi:\discretionary{}{}{}\begingroup
  \urlstyle{rm}\Url}\fi

\bibitem[{Beheshti and Dahleh(2005)}]{paper:beheshti2005new}
Beheshti, S. and Dahleh, M.A. (2005).
\newblock A new information-theoretic approach to signal denoising and best
  basis selection.
\newblock \emph{IEEE Transactions on Signal Processing}, 53(10), 3613--3624.

\bibitem[{De~Moor(2019)}]{de2019least}
De~Moor, B. (2019).
\newblock Least squares realization of lti models is an eigenvalue problem.
\newblock In \emph{2019 18th European Control Conference (ECC)}, 2270--2275.
  IEEE.

\bibitem[{Doclo and Moonen(2003)}]{paper:doclo2003robust}
Doclo, S. and Moonen, M. (2003).
\newblock Robust adaptive time delay estimation for speaker localization in
  noisy and reverberant acoustic environments.
\newblock \emph{EURASIP Journal on Advances in Signal Processing}, 2003(11),
  495250.

\bibitem[{Elliott and Nelson(1993)}]{paper:elliott1993active}
Elliott, S.J. and Nelson, P.A. (1993).
\newblock Active noise control.
\newblock \emph{IEEE signal processing magazine}, 10(4), 12--35.

\bibitem[{Georgiou et~al.(1992)Georgiou, Shankwitz, and
  Smith}]{georgiou1992identification}
Georgiou, T.T., Shankwitz, C.R., and Smith, M.C. (1992).
\newblock Identification of linear systems: A graph point of view.
\newblock In \emph{1992 American Control Conference}, 307--312. IEEE.

\bibitem[{Haykin(2008)}]{book:haykin2008adaptive}
Haykin, S.S. (2008).
\newblock \emph{Adaptive filter theory}.
\newblock Pearson Education India.

\bibitem[{Joliffe(2002)}]{book:joliffe}
Joliffe, I. (2002).
\newblock \emph{Principal component analysis}.
\newblock Statistics. Springer-Verlag, New York, USA.

\bibitem[{Ljung(1999)}]{book:ljung}
Ljung, L. (1999).
\newblock \emph{System Identification - A Theory for the User}.
\newblock Prentice Hall International, Upper Saddle River, NJ, USA.

\bibitem[{Maleki et~al.(2014)Maleki, Shang, and Rapisarda}]{soton368264}
Maleki, S., Shang, Z., and Rapisarda, P. (2014).
\newblock A geometric approach to fault identification in linear repetitive
  processes.
\newblock In \emph{The 21st International Symposium on Mathematical Theory of
  Networks and Systems (MTNS 2014)}.
\newblock \urlprefix\url{https://eprints.soton.ac.uk/368264/}.

\bibitem[{Maurya et~al.(2016)Maurya, Tangirala, and
  Narasimhan}]{paper:dycops_dipca}
Maurya, D., Tangirala, A.K., and Narasimhan, S. (2016).
\newblock Identification of linear dynamic systems using dynamic iterative
  principal component analysis.
\newblock \emph{IFAC-PapersOnLine}, 49(7), 1014--1019.

\bibitem[{Maurya et~al.(2018)Maurya, Tangirala, and
  Narasimhan}]{paper:iecr_dipca}
Maurya, D., Tangirala, A.K., and Narasimhan, S. (2018).
\newblock Identification of errors-in-variables models using dynamic iterative
  principal component analysis.
\newblock \emph{Industrial \& Engineering Chemistry Research}, 57(35),
  11939--11954.

\bibitem[{Moler and Stewart(1973)}]{paper:qz_moler}
Moler, C.B. and Stewart, G.W. (1973).
\newblock An algorithm for generalized matrix eigenvalue problems.
\newblock \emph{SIAM Journal on Numerical Analysis}, 10(2), 241--256.

\bibitem[{Moore et~al.(2013)Moore, Brookes, and Naylor}]{paper:moore2013room}
Moore, A.H., Brookes, M., and Naylor, P.A. (2013).
\newblock Room geometry estimation from a single channel acoustic impulse
  response.
\newblock In \emph{Signal Processing Conference (EUSIPCO), 2013 Proceedings of
  the 21st European}, 1--5. IEEE.

\bibitem[{Morikawa and Fujisaki(1984)}]{paper:morikawa1984system}
Morikawa, H. and Fujisaki, H. (1984).
\newblock System identification of the speech production process based on a
  state-space representation.
\newblock \emph{IEEE transactions on acoustics, speech, and signal processing},
  32(2), 252--262.

\bibitem[{Narasimhan and Shah(2008)}]{paper:ipca}
Narasimhan, S. and Shah, S.L. (2008).
\newblock Model identification and error covariance matrix estimation from
  noisy data using pca.
\newblock \emph{Control Engineering Practice}, 16(1), 146--155.

\bibitem[{Oppenheim(1999)}]{book:oppenheim1999discrete}
Oppenheim, A.V. (1999).
\newblock \emph{Discrete-time signal processing}.
\newblock Pearson Education India.

\bibitem[{Perepu and Tangirala(2015)}]{paper:satheesh_CS}
Perepu, S.K. and Tangirala, A.K. (2015).
\newblock Identification of equation error models from small samples using
  compressed sensing techniques.
\newblock \emph{IFAC-PapersOnLine}, 48(8), 795--800.

\bibitem[{Rao(1964)}]{paper:rao1964pca}
Rao, C.R. (1964).
\newblock The use and interpretation of principal component analysis in applied
  research.
\newblock \emph{Sankhy{\=a}: The Indian Journal of Statistics, Series A},
  329--358.

\bibitem[{Rapisarda(2018)}]{soton425875}
Rapisarda, P. (2018).
\newblock On the identification of self-adjoint linear time-varying state
  models.
\newblock \emph{IFAC-PapersOnLine}, 51(15), 251--256.

\bibitem[{Sepulchre et~al.(2012)Sepulchre, Jankovic, and
  Kokotovic}]{sepulchre2012constructive}
Sepulchre, R., Jankovic, M., and Kokotovic, P.V. (2012).
\newblock \emph{Constructive nonlinear control}.
\newblock Springer Science \& Business Media.

\bibitem[{Shumway and Stoffer(2000)}]{shumway2000time}
Shumway, R.H. and Stoffer, D.S. (2000).
\newblock Time series analysis and its applications.
\newblock \emph{Studies In Informatics And Control}, 9(4), 375--376.

\bibitem[{Tangirala(2014)}]{book:akt_sysid}
Tangirala, A.K. (2014).
\newblock \emph{Principles of System Identification: Theory and Practice}.
\newblock CRC Press, Taylor \& Francis Group, Boca Raton, FL, USA.

\bibitem[{Vermeersch and De~Moor(2019)}]{vermeersch2019globally}
Vermeersch, C. and De~Moor, B. (2019).
\newblock Globally optimal least-squares arma model identification is an
  eigenvalue problem.
\newblock \emph{IEEE Control Systems Letters}, 3(4), 1062--1067.

\bibitem[{Wax(1988)}]{paper:wax1988order}
Wax, M. (1988).
\newblock Order selection for ar models by predictive least squares.
\newblock \emph{IEEE Transactions on Acoustics, Speech, and Signal Processing},
  36(4), 581--588.

\bibitem[{Xu et~al.(1995)Xu, Liu, Tong, and Kailath}]{paper:xu1995least}
Xu, G., Liu, H., Tong, L., and Kailath, T. (1995).
\newblock A least-squares approach to blind channel identification.
\newblock \emph{IEEE Transactions on signal processing}, 43(12), 2982--2993.

\bibitem[{Xue et~al.(2017)Xue, Brookes, and Naylor}]{paper:xue2017frequency}
Xue, W., Brookes, M., and Naylor, P.A. (2017).
\newblock Frequency-domain under-modelled blind system identification based on
  cross power spectrum and sparsity regularization.
\newblock In \emph{Acoustics, Speech and Signal Processing (ICASSP), 2017 IEEE
  International Conference on}, 591--595. IEEE.

\bibitem[{Zheng and Ohta(2018)}]{zheng2018positive}
Zheng, M. and Ohta, Y. (2018).
\newblock Positive fir system identification using maximum entropy prior.
\newblock \emph{IFAC-PapersOnLine}, 51(15), 7--12.

\end{thebibliography}
                                                     % with bibtex (preferred)

\appendix
\section{Appendix: QZ Algorithm}
\label{app:appendix}
The QZ algorithm  \citep{paper:qz_moler} is a numerical method to solve the generalized eigenvalue problem, $\mathbf{Ax} = \lambda \mathbf{ B x}$. The eigenvalues and eigenvectors can be easily obtained for invertible $\mathbf{ B}$ but QZ algorithm solves the above equation without computation of $\mathbf{ B}^{-1}$. 

The key idea is to simultaneously transform $\mathbf{ A}$ and $\mathbf{ B}$ into generalized Schur form by using similarity transformations. This may be stated as we intend to find orthogonal matrices $\mathbf{ Q}$ and $ \mathbf{ Z}$
\begin{align}
\mathbf{ Q^T (A - \lambda B) Z = T - \lambda S}
\end{align}
% More specifically,  $A$ is reduced to upper Hessenberg form and $B$ is reduced to upper triangular form . 
where  $\mathbf{ T}$ is of quasi-upper triangular form and  $\mathbf{ S}$ is of upper triangular shape.
Eigenvalues and eigenvectors can be computed from the diagonals of triangular form. 
% We encourage readers to refer \cite{qz_1973} for more insights.
% Eigenvectors can be computed as the eigenvectors of the triangular problem and then transformed back with $Z$ to the eigenvectors of the original problem.
                                                                      % in the appendices.
\end{document}